\documentclass[oneside]{article}

\usepackage{PRIMEarxiv}

\usepackage[utf8]{inputenc} 
\usepackage[T1]{fontenc}    
\usepackage{hyperref}       
\usepackage{url}            
\usepackage{booktabs}       
\usepackage{amsfonts}       
\usepackage{nicefrac}       
\usepackage{microtype}      
\usepackage{lipsum}
\usepackage{fancyhdr}       
\usepackage{graphicx}       
\graphicspath{{media/}}     

\usepackage{comment}
\usepackage[table,xcdraw]{xcolor}

\usepackage{amsmath}
\newcommand\norm[1]{\left\lVert#1\right\rVert}

\usepackage{adjustbox}

\usepackage{algorithm}      
\usepackage[noend]{algpseudocode}   

\usepackage{listings}       
\usepackage{color}
\definecolor{lightgray}{rgb}{.9,.9,.9}
\definecolor{darkgray}{rgb}{.4,.4,.4}
\definecolor{key_purple}{RGB}{194,130,180}
\definecolor{olivegreen}{RGB}{109,113,46}
\definecolor{navy_blue}{RGB}{81, 106, 174}
\definecolor{base_text_blue}{RGB}{173, 216, 250}
\definecolor{base_text_yellow}{RGB}{204,178,48}
\definecolor{water_green}{RGB}{35, 174, 186}
\definecolor{light_yellow}{RGB}{255, 255, 167}
\definecolor{base_black}{RGB}{35, 35, 35}
\definecolor{brick_red}{RGB}{ 182,123,90}

\lstdefinelanguage{Python}{
    keywords=[1]{
        try, except, for, if, else, return, in
    },
    keywordstyle=[1]\color{key_purple}\bfseries,
    morecomment=[l]{\#},
    morecomment=[s]{/*}{*/},
    morestring=[b]',
    morestring=[b]",
    morekeywords=[2]{
        class, def, 
    },
    keywordstyle=[2]\color{navy_blue}\bfseries,
    morekeywords=[3]{
        np, df, plt, tf, enumerate, int, go, 
    },
    keywordstyle=[3]\color{water_green}\bfseries,
    morekeywords=[4]{
        print, 
        create_roi, get_Energy_data
    },
    keywordstyle=[4]\color{light_yellow}\bfseries,
    morekeywords=[5]{
        sum, abs, where, clip, astype, max, Figure, moveaxis, Image, =, !, 
    },
    keywordstyle=[5]\color{white}\bfseries,
    identifierstyle=\color{base_text_blue},
    commentstyle=\color{olivegreen}\ttfamily,
    stringstyle=\color{brick_red}\ttfamily,
    sensitive=true
}

\newcommand{\lea}{\leftarrow}

\newcommand{\np}{\texttt{NumPy}}

\title{ganX - generate artificially new XRF \\ \textit{a python library to generate MA-XRF raw data out of RGB images}
}

\author{
  \href{https://orcid.org/0000-0001-7225-3355}{\includegraphics[scale=0.06]{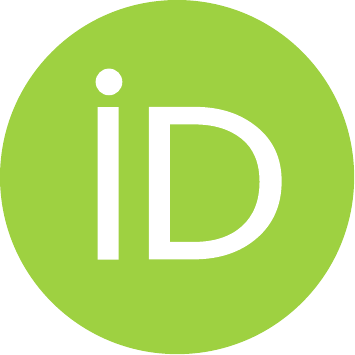} }Alessandro Bombini \\
  Istituto Nazionale di Fisica Nucleare \\
  via B. Rossi 1, 50019, \\
  Sesto Fiorentino (FI)\\
   \texttt{bombini@fi.infn.it}
}

\begin{document}
\maketitle

\begin{abstract}
In this paper we present the first version of \textit{ganX - generate artificially new XRF}, a Python library to generate X-ray fluorescence Macro maps (MA-XRF) from a coloured RGB image. To do that, a Monte Carlo method is used, where each MA-XRF pixel signal is sampled out of an XRF signal probability function. Such probability function is computed using a database of couples (pigment characteristic XRF signal, RGB), by a weighted sum of such pigment XRF signal by proximity of the image RGB to the pigment characteristic RGB. The library is released to PyPi and the code is available open source on GitHub. 
\end{abstract}


\section{Introduction}

In the last decade, we have witnessed a truly remarkable rise of Artificial Intelligence, Statistical and Deep learning methods (for a non exhaustive list of papers on the history of deep learning, see \cite{https://doi.org/10.48550/arxiv.1702.07800, Goodfellow-et-al-2016, d2ai/arxiv.2106.11342}, and references therein).

Inspired by the incredible results obtained thanks to the application of such methods to scientific problems, its adoption in the field of nuclear imaging applied to Cultural Heritage (CH) has begun (see, e.g., \cite{Kleynhans2020,5967899,8432512,8667664,666999xrfdl}, especially the nice overview \cite{s23052419}, and, of course, the references therein), also in the field of X-ray fluorescence Macro mapping (MA-XRF) \cite{DBLP:journals/corr/abs-1812-10836, BombiniICIAP2021, BombiniICCSA2022, D2JA00114D, Jones2022, arxiv.2207.12651}. In MA-XRF, the imaging apparatus produces a data cube which, for each pixel, is formed by a spectrum containing fluorescence lines associated with the element composition of the pigment present in the pictorial layers.

MA-XRF data cubes offer an ideal framework for application of unsupervised statistical learning methods \cite{D2JA00114D}, due to the huge number of pixel XRF histogram w.r.t.~the relatively small number of employed pigment palettes. 

Unfortunately, in the realm of supervised statistical (deep) learning applied to CH-based analysis, the situation is flipped \cite{Jones2022}, since the data cube production is slow, obtaining a small dataset for the complexity of the various task at hand (like automatic pigment identification \cite{Jones2022,arxiv.2207.12651}, element recognition \cite{9511278}, and even colour association \cite{BombiniICIAP2021, BombiniICCSA2022}).   

This justifies the emphasis put on the creation of ad hoc \textit{synthetic} MA-XRF dataset \cite{Jones2022}. In this brief paper, we present \texttt{ganX} \cite{Bombini_ganX_-_generate_2023}, a Python library for creating synthetic dataset of MA-XRF images, starting from 
\begin{enumerate}
    \item A (set of) RGB image(s); and
    \item A Database of pigments;
\end{enumerate}
Such database must have
\begin{itemize}
    \item[a] A characteristic XRF signal;
    \item[b] A characteristic, 8-bit, RGB colour. 
\end{itemize}

\texttt{ganX} will use (b) to associate pigment(s) to RGB pixels, and (a) as a probability distribution to generate the MA-XRF data cube via Monte Carlo methods. 

At the day of writing, the package is available open access on the Python Package index (PyPi)\footnote{\url{https://pypi.org/project/ganx/}.}. The package code is open source on GitHub \cite{Bombini_ganX_-_generate_2023}, where its documentation is available. 

\section{Main idea}

Inspired by the work done in \cite{Jones2022}, where the authors generate a synthetic dataset of single-point XRF histogram by using the Fundamental Parameters (FP) method, a physical model based on Sherman's equations for generating XRF spectra of a sample of known composition \cite{sherman1955theoretical}, we started the development of a simple python package capable of creating MA-XRF data cubes starting from a RGB image and a database of pigments' XRF signal. 

The data generation algorithm comprises two main parts:
\begin{enumerate}
    \item A RGB clustering of the input image(s), performed with an iterative K-Means, to reduce the colour noise;
    \item A Monte Carlo method to generate a XRF histogram of counts for each pixel, after having unsupervisedly associated a (set of) pigment(s) to each RGB pixel.
\end{enumerate}

To perform step 2.~above, we compute (pixel-by-pixel) a set of similarity measures in RGB spaces, between the (clustered) RGB and the database indexed pigments' characteristic RGB; then, an hard thresholding is performed, to (unsupervisedly) reduce the number of pigments. 

The full pseudocode of the MA-XRF data generation is
\begin{algorithm}[H]
\caption{Generate MA-XRF}\label{alg:GenerateMA-XRF}
\begin{algorithmic}[1]
\Require \textsc{RGBclustering}, \textsc{distrToXrf}, \textsc{colorSimilarity} \Comment{Import relevant methods}

\Procedure{generateXRF}{pigments\_dict\_data = $\mathcal{D}_{{color}, {hist}}$, rgb\_img = $R$}
\State $R \leftarrow$ \textsc{RGBclustering}($R$)
\State init $X$; \Comment{Initialise empy XRF data cube}
\For{$r\in R$} \Comment{Iterate over clusters}
    \State init $d$; \Comment{Initialise empty distribution}
    \For{$(c, h) \in \mathcal{D}$} \Comment{Iterate over database}
        \State $\alpha \leftarrow$  \textsc{colorSimilarity}($c, r$) 
        \If{$\alpha \ge \alpha_{th}$} 
            \State $d \leftarrow d + \alpha \cdot h$          \Comment{Sum thresholded distribution}   
        \EndIf
    \EndFor

    \State $d \leftarrow d /\norm{d}$ \Comment{Normalise distribution}
    \State $X[\textrm{idx}(X)== r] \leftarrow $ \textsc{distrToXrf}($d$) \Comment{Use Monte Carlo method only on the Cluster's pixel}
\EndFor 

\State \textbf{return} $X$

\EndProcedure
\end{algorithmic}
\end{algorithm}

The actual code implementation in Python will leverage the Numpy \cite{van_der_Walt_2011_Numpy,harris2020array} \textit{vectorization} to speed up the computation. 

The details of the (pseudo)function \textsc{RGBclustering, distrToXrf, colorSimilarity}, are reported in the following subsections.

\paragraph{Note on \cite{Jones2022}:} Notice that it is possible to use the FP approach of \cite{Jones2022} to build the characteristic pigments' XRF signal; this implies that it is straightforward to use the \texttt{ganX} library to leverage the synthetic data generation of \cite{Jones2022} from single-point XRF to MA-XRF data cubes. 

On the other hand, \texttt{ganX} can be used out of \textit{real} data, comprising measurement whose source origin is complicated enough that no analythic method (such as the FP method) can be used to generate the XRF signal. This latter situation is quite common in the field of Cultural Heritage applications. 

\paragraph{Additional note:} Notice also that the spectral signal is not limited to be an XRF characteristic signal; it is possible to generate data cubes out of other spectroscopy techniques, not even in the X-Ray range, such as Fourier Transform InfraRed spectroscopy (FTIR), HyperSpectral Imaging (HSI), Particle Induced X-Ray Emission (PIXE), etc., just to name a few.

\subsection{Iterative KMeans clustering}

Since the algorithm to generate the MA-XRF data cube relies on the "similarity" between a (set of) reference RGB(s) to the input RGB, an RGB segmentation of the input image is performed to reduce the variance in RGB space, and thus, the noise in the similarity computation. 

To perform the clustering, K-Means algorithm \cite{kmeans1967originalpaper} was used, in its \textit{scikit-learn} implementation \cite{scikit-learn}. One of the main disadvantages of K-Means is that the number of cluster is an \textit{a priori} parameter. Another issue is its long convergence time. 

To address both of those issue, we either
\begin{itemize}
    \item defined an iterative procedure for evaluating K-Means with different number of centroids;
    \item used the MiniBatch K-Means algorithm of scikit-learn;
\end{itemize}
The iteration is performed from a starting value $n_0$ of clusters, which is the  central value of the iteration, which is performed from $n_i = n_0 - \delta N$ to $n_f = n_0 + \delta N$; $\delta N$ is an additional parameters (defaults to 3). Also, there is a $n_{patience}$ additional argument (i.e.,~if there is no improvement after $n_{patience}$ iterations, the cycle is broken). 

To unsupervisedly compute the performance of the KMeans algorithm at each step, we use the \textit{Silhouette Score} \cite{ROUSSEEUW198753, silhouettescore1990}\footnote{The Silhouette Score $s$ is computed as the mean Silhouette Coefficient $s_i$ of all samples, i.e. 
\begin{equation}
    s = \max_{I} \left[ \frac{1}{\# C_I} \sum_{i\in C_i} s_i \right] \,,
\end{equation}
where $s_i$ is computed as
\begin{equation}
    s_i = \frac{b_i - a_i}{\max_i (a_i, b_i)} \,,  
\end{equation}
and where, denoting the centroid of the $I$-th cluster $C_I$ with $\mu_{C_I}$, 
\begin{equation}
    a_i = \textrm{dist}(i, \mu_{C_I})\,, \quad b(i) = \min_{C_J \neq C_I} \textrm{dist}(i, C_J) \, . 
\end{equation}
}

The pseudocode implementation of \textsc{RGBclustering} is:

\begin{algorithm}[H]
\caption{IterativeKMeans}\label{alg:IterativeKMeans}
\begin{algorithmic}[1]

\Require \textsc{MiniBatchKMeans}, \textsc{silhouette\_score} \Comment{Those are standard sklearn class and methods}

\Procedure{IterativeKMeans}{rgb\_img = $R$, $n_0$}
\State $n_i = n_0 - \delta N$
\State $n_f = n_0 + \delta N$
\State init $C_f$;
\State init $s_{old}\lea -1$;
\State init count;
\For{$n \lea n_i, n_f $}
    \State $C \lea $ \textsc{MiniBatchKMeans}($n$).\textsc{fit\_predict}($R$) \Comment{Instantiate, Fit and Predict}
    \State $s\lea $ \textsc{silhouette\_score} ($C$)
    \If{$s>s_{old}$}
        \State $C_f\lea C$  \Comment{Update the clustering result if this iteration has higher score}
    \Else
        \State count++
        \If{count $\ge n_p$}
            \State \textbf{break}; \Comment{Break loop if we hit the patience limit}
        \EndIf
    \EndIf
    
\EndFor 

\State \textbf{return} $C_f$

\EndProcedure
\end{algorithmic}
\end{algorithm}

After that, it is possible to compute the average colour for each cluster and build the clustered image:

\begin{algorithm}[H]
\caption{Clustering RGB images}\label{alg:RGBclustering}
\begin{algorithmic}[1]

\Require \textsc{IterativeKMeans}

\Procedure{RGBclustering}{rgb\_img = $R$, $n_0$}
\State $C \lea $ \textsc{IterativeKMeans}($R$, $n_0$)
\State init $R_c$, $r$, $M$;
\For{$C_I \in C $}
    \State $r_I \lea \frac{1}{\# C_i} \sum_{i\in C_I} R_i$ \Comment{Compute the average RGB in the cluster}
    \State $M_I \lea $ Bool($R[C_I]$) \Comment{Compute the boolean mask of the cluster}
    \State $R_c \lea R_c + r_I \cdot M_I$ \Comment{Add the cluster RGB image to the full clustered image}
\EndFor 
\State \textbf{return} $R_c$, $M$, $r$
\EndProcedure
\end{algorithmic}
\end{algorithm}

\subsection{Generating MA-XRF with Monte Carlo methods}

To generate the MA-XRF, we use a Monte Carlo method; to do so, we have to assign a probability distribution of XRF signals to an RGB colour. In order to do that, we compute a weighted thresholded sum of indexed XRF signals, where the weights are a similarity measure, as explained in Algorithm \ref{alg:GenerateMA-XRF}. In particular, the \textsc{colorSimilarity} method is simply a Delta CIEDE 2000 $\delta_{2000}$ function \cite{ciede2000paper,ciede2000implementation} in the scikit-image package \cite{scikit-image}, so that
\begin{equation}
    \textsc{colorSimilarity} (x,y) = \frac{100 - \delta_{2000}(x, y)}{100} \,.
\end{equation}
Notice that, since $\delta_{2000}(x, y) \in [0,100] \; \forall x,y$, we have that $\textsc{colorSimilarity} (x,y) \in [0,1]$, where low values represent dissimilar colours, while high values represent perceptually similar colours.

In the class exposing those distance measuring methods, we have implemented also Delta CIEDE 1994 function \cite{ciede1994paper}, delta76 function \cite{ciede1976paper}, as well as the cosine similarity. 

Also, to generate the MA-XRF, we use a Monte Carlo method with the aforementioned discrete probability function. The na\"ive, slow implementation of such method works pixel-by-pixel:

\begin{algorithm}[H]
\caption{Slow Monte Carlo}\label{alg:slowDistrToXrf}
\begin{algorithmic}[1]
\Require \textsc{NumPy} \textbf{as} \texttt{np}
\Procedure{slowDistrToXrf}{num\_of\_counts = $N$, distr = $h$, size = $s$}
    \State init XRF; 
    \For{$i\lea 0, s$}
        \State XRF$[i] \lea$ slowMonteCarlo($N$, $h$)
    \EndFor

    \State \textbf{return} XRF
\EndProcedure
\State 
\Procedure{slowMonteCarlo}{num\_of\_counts = $N$, distr = $h$}
    \State Hist1D $\lea$ \texttt{np.random.choice}( 
            \texttt{np.arange}(0, $h$.\texttt{size}),
            size = $N$, 
            p = $h$
    ) \Comment{Perform the Monte Carlo}
    \State $X\lea$ \texttt{np.bincount}( Hist1D, minlength = $h$.\texttt{size} ) \Comment{Get the XRF}
    \State \textbf{return} X
\EndProcedure
\end{algorithmic}
\end{algorithm}

The faster implementation of the Monte Carlo methods relies on NumPy vectorisation, and the usage of NumPy Histogram2D generator, namely the \texttt{ np.random.default\_rng().choice} method\footnote{See NumPy Random documentation page \url{https://numpy.org/doc/stable/reference/random/generator.html}. }; in order to do so, we have to define a two-dimensional bincount, which, at the date of writing, is unavailable in the NumPy package:

\begin{algorithm}[H]
\caption{Monte Carlo}\label{alg:DistrToXrf}
\begin{algorithmic}[1]
\Require \textsc{NumPy} \textbf{as} \texttt{np}
\Procedure{DistrToXrf}{num\_of\_counts = $N$, distr = $h$, size = $s$}
    \State rng $\lea$ \texttt{np.random.default\_rng}() 
    \State Hist2D$\lea$ rng.\texttt{choice}(
        $h$.\texttt{size},
        size = ($s$, $N$),
        p = $h$,
        axis = -1
    )
    \State XRF $\lea$ \textsc{bincount2d}(
        arr = Hist2D, 
        bins = $h$.\texttt{size}
    )
    \State \textbf{return} XRF
    
\EndProcedure
\State 
\Procedure{bincount2d}{arr, bins}
    \If{bins is \texttt{None}}
        \State bins $\lea \max$(arr) + 1  \Comment{Get the bins as the max in arr}
    \EndIf
    \State init count;
    \State indexing $\lea$ (\texttt{np.ones\_like}(arr).T $\cdot$ \texttt{np.arange}(len(arr))).T \Comment{Get the indexes}
    \State \texttt{np.add.at}(count, (indexing, arr), 1) \Comment{Insert counts}
    \State \textbf{return} count
\EndProcedure
\end{algorithmic}
\end{algorithm}

The code here is highly \textit{pythonic}, and relies heavily on the \texttt{NumPy} grammar; the key difference between Algorithm \ref{alg:slowDistrToXrf} and Algorithm \ref{alg:DistrToXrf} is the output shape of the Monte Carlo generator.

In the first case (Algorithm \ref{alg:slowDistrToXrf}), the method \texttt{np.random.choice} generates $N$ random extraction of counts in the range (0, $h$.\texttt{size}) (i.e.,~the X-ray Energy bins), with each (discrete) bin extraction with the (discrete) probability given by $h$ (i.e.,~the passed XRF distribution). The resulting array is thus passed to a bin count function \texttt{np.bincount}, with array size fixed to be $h$.\texttt{size} (i.e.,~the same of the passed XRF distribution). This algorithm is slow because we need a slow, non-pythonic for loop.

In the second case (Algorithm \ref{alg:DistrToXrf}), the class \texttt{np.random.default\_rng}  method \texttt{choice} allows for a 2D array generation, with shape $(s, N)$, i.e.~ $N$ extractions for each of the $s$ lines\footnote{Notice that, here, $s$ is the product of lines and rows of the MA-XRF.}. This is way faster than the for loop due to the fact that \texttt{NumPy} arrays are densely packed arrays of homogeneous type (thus get the benefits of \textit{locality of reference}), and also because these element-wise operations are gouped together (ofter referred as \textit{vectorisation}) \cite{van_der_Walt_2011_Numpy}, and are implemented in the strongly-typed language \texttt{C}. Since, at the day of writing, there is no higher-dimensional bin count method natively implemented in \np, we have to write it down one, which heavily employs \textit{vectorisation} to speed up computation. Fortunately, it is possible to define such a method\footnote{User \texttt{winwin} has posted on StackOverflow a very nice implementation of such function; See \url{https://stackoverflow.com/questions/19201972/can-numpy-bincount-work-with-2d-arrays}.}, that is the \textsc{bincount2d} function reported in Algorithm \ref{alg:DistrToXrf}; the \textsc{bincount2d} function works in two steps:
\begin{enumerate}
    \item build a two-dimensional \np~ array $\texttt{indexing}$ of shape ($s$, $N$) (i.e.~the same of the input variable $\texttt{arr}$), where the entry $\texttt{indexing}_{\text{row},  \text{col}}$  is equal to $\text{row}$ (i.e.~a row of 0s, a row of 1s, etc.) 
    \item use the vectorised \texttt{np.add.at} method to add at the \texttt{count} \np~array the appropriate number of counts at the appropriate index.
\end{enumerate}


\section{A simple usage example}\label{sec:simple_example}

We have built a small pigment dataset comprising 9 pigments:
\begin{enumerate}
    \item Lead Red
    \item Vermillion
    \item Organic Red Dye
    \item Lead Tin Yellow
    \item Ultramarine
    \item Azurite
    \item Indigo
    \item Malachite
    \item Chalk white
\end{enumerate}
Those pigments in the list were chosen either by data availability\footnote{To populate the database of the example reported in thi section, fro the XRF signals, we used the XRF pigment signal data from \cite{cortea2023}.}, as well as similarity with the pigment palette of the work \cite{heritage2020103}, where were conducted a set of non-invasive analysis on illuminated manuscripts. 

\begin{figure}
    \centering
    \includegraphics[width = 0.8\textwidth]{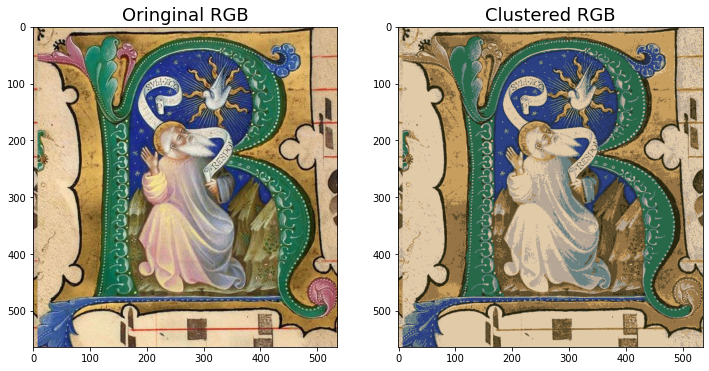}
    \caption{On the left: RGB image of Antiphonary M, folio 32; from \cite{heritage2020103}. On the right: clustered RGB with Iterative K-Means. Best value in the range $(8,12)$ at $K=11$.}
    \label{fig:AntM_f32_RGB}
\end{figure}

We had access to the raw data of a single MA-XRF map obtained in \cite{heritage2020103}, the map of Antiphonary M, folio 32\footnote{ Antiphonary M belongs to the collection of the Abbey of San Giorgio Maggiore, Venice.} (see Figure \ref{fig:AntM_f32_RGB}).

\begin{figure}[h]
    \centering
    \includegraphics[width=0.97\textwidth]{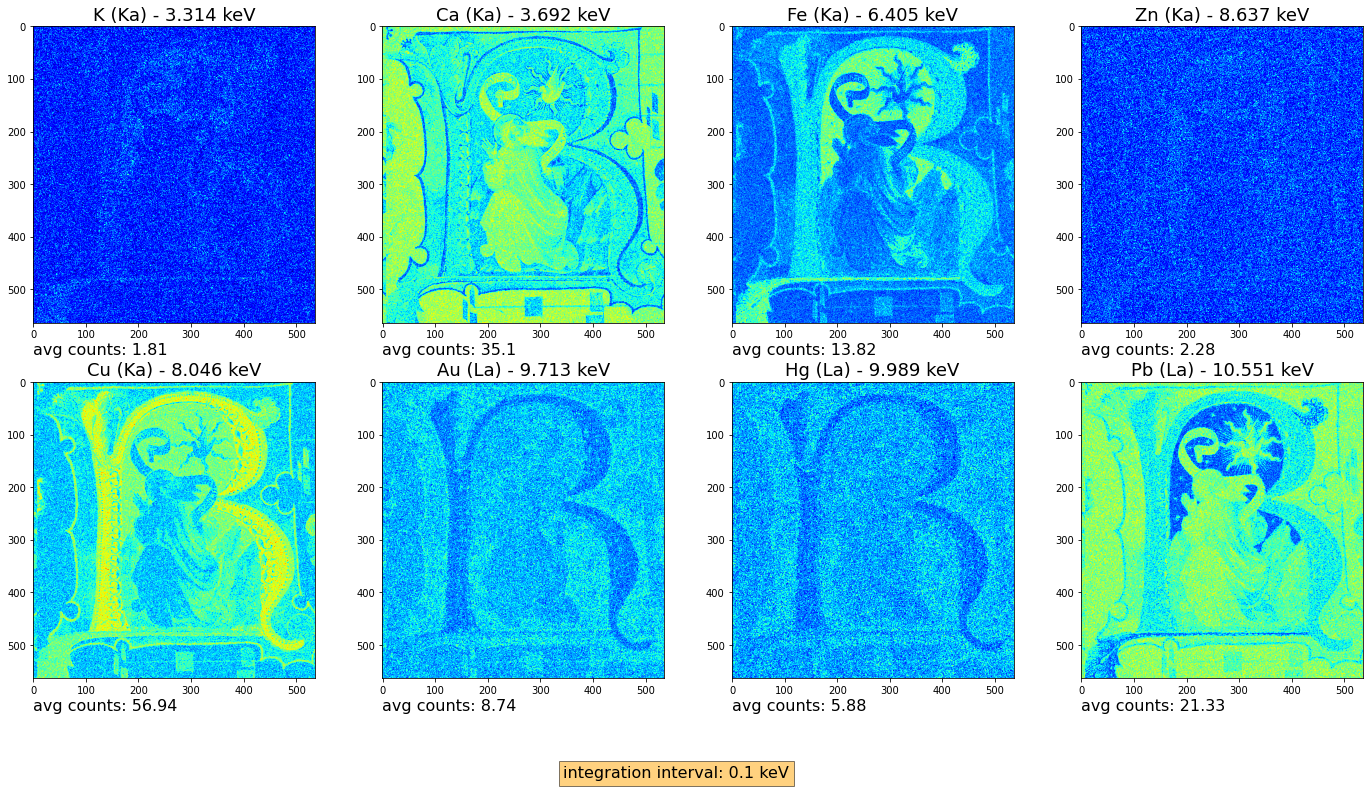}
    \caption{Examples of Elemental maps of the generated XRF with threshold $\alpha = 0.2$.}
    \label{fig:fake_elements}
\end{figure}

\begin{figure}[h]
    \centering
    \includegraphics[width=0.97\textwidth]{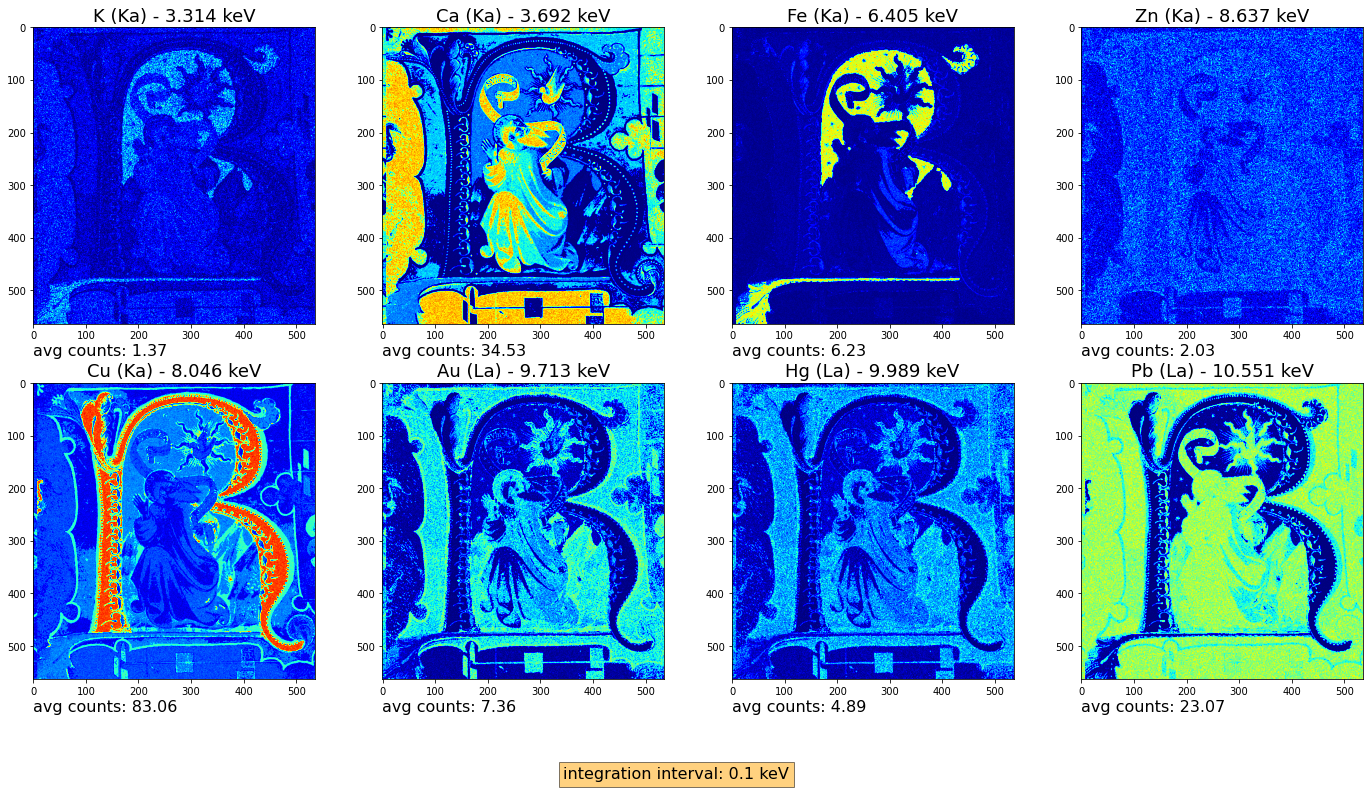}
    \caption{Examples of Elemental maps of the generated XRF with threshold $\alpha = 0.6$.}
    \label{fig:fake_elements_06}
\end{figure}

\begin{figure}[h]
    \centering
    \includegraphics[width=0.97\textwidth]{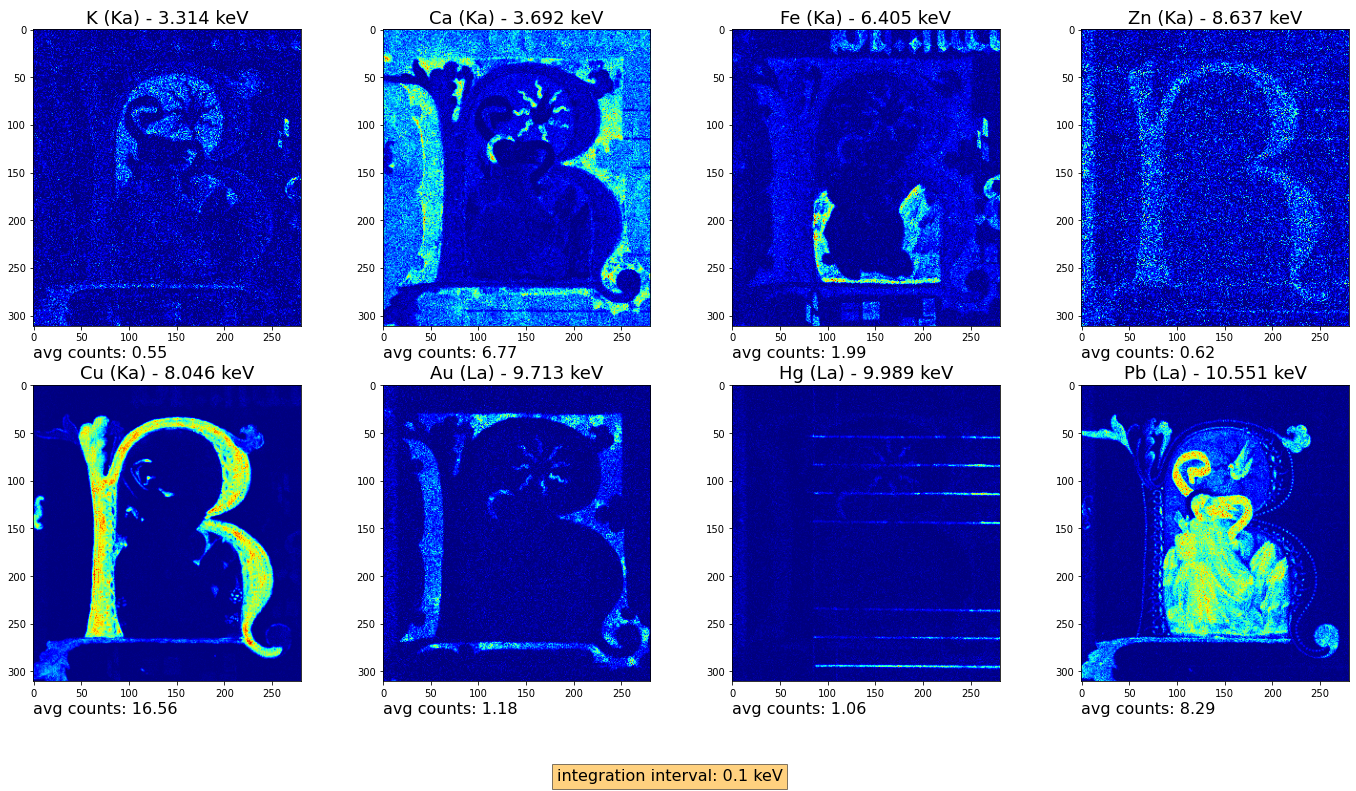}
    \caption{Examples of Elemental maps of the true XRF}
    \label{fig:true_elements}
\end{figure}

We have thus generated the synthetic MA-XRF using the \texttt{XRFGenerator} class of \texttt{ganX}, and we stored it as an HDF5 file. To do so, w have used \texttt{\_num\_of\_counts}$ = 400$, $\texttt{N\_start} = 10$, $\texttt{delta\_N} = 2$. $\texttt{N\_patience} = -1$, $\texttt{score\_batch} = 1024$ as parameters. As an example, we have performed the generation twice: the first, with $\texttt{generation\_threshold}=0.2$; the second,  with $\texttt{generation\_threshold}=0.6$. The choice was mostly random, with the intent to show how different thresholds impact the presence of pigments' contribution in pixel counts. This is visually evident in Figures \ref{fig:fake_elements}, \ref{fig:fake_elements_06}.

\begin{figure}[h]
    \centering
    \includegraphics[width=0.97\textwidth]{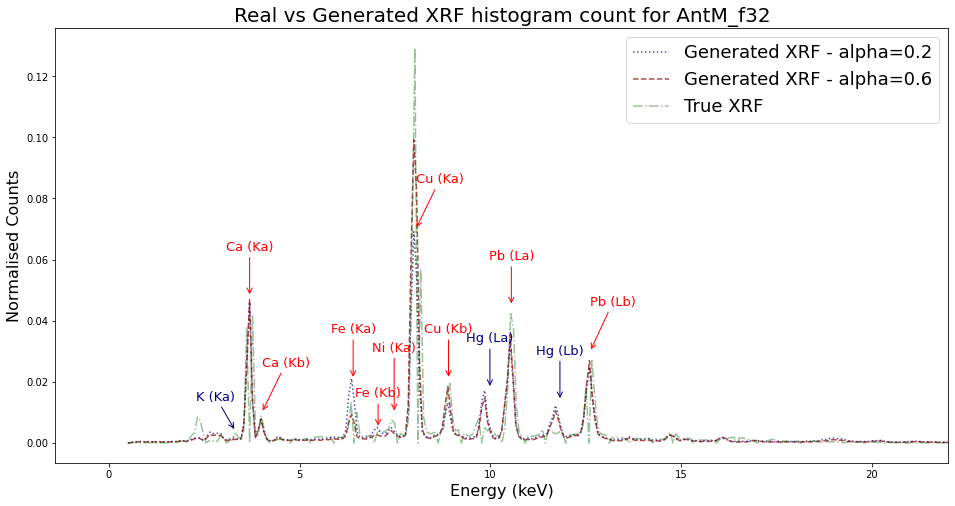}
    \caption{Comparison of the total XRF counts}
    \label{fig:hists}
\end{figure}

To visually inspect the generated result, we computed 8 elemental maps: K (K$\alpha$), Ca (K$\alpha$), Fe (K$\alpha$), Zn (K$\alpha$), Cu (K$\alpha$), Au (L$\alpha$), Hg (L$\alpha$), Pb (L$\alpha$). The results are reported in Figures \ref{fig:fake_elements}, \ref{fig:fake_elements_06}, \ref{fig:true_elements}.


It is possible to give a numerical hint on the similarity between fake and true XRF in some different ways. We start by inspecting the integrated image histogram, reported in Figure \ref{fig:hists}. There are reported the generated XRF with threshold $\alpha=0.2$ histogram in dotted blue, the generated XRF with threshold $\alpha=0.6$ histogram in dashed red, and the true XRF histogram in dash-dotted green. The results for various metrics are reported in Table \ref{tab:hist_res}. For a brief description of the metrics, see Appendix \ref{app:metrics}

\begin{table}[h]
\caption{Numerical results of comparison between the fake and true XRF histograms reported in Figure \ref{fig:hists}.}\label{tab:hist_res}
\centering
\begin{tabular}{l|cc}
                                                                 & \cellcolor[HTML]{FFCCC9}$\alpha=0.2$ & \cellcolor[HTML]{FFCCC9}$\alpha=0.6$ \\ \hline
\cellcolor[HTML]{FFFFC7}{\color[HTML]{333333} Cosine Distance}   & 0.2191                               & 0.1977                               \\
\cellcolor[HTML]{FFFFC7}{\color[HTML]{333333} Jensen-Shannon}    & 0.3465                               & 0.3363                               \\
\cellcolor[HTML]{FFFFC7}{\color[HTML]{333333} Chebyshev}         & 0.0602                               & 0.0817                               \\ 
\cellcolor[HTML]{FFFFC7}Correlation                              & 0.2343                               & 0.2105                               \\ 
\cellcolor[HTML]{FFFFC7}Bray-Curtis                              & 0.3528                               & 0.3203                               \\
\cellcolor[HTML]{FFFFC7}Chi-square                                & 0.3996                               & 0.3706                               \\ \hline
\cellcolor[HTML]{ECF4FF}Bhattacharyya                            & 0.8529                               & 0.8589                              
\end{tabular}
\end{table}

The yellow background metrics represent distances (i.e.~the lower, the better), while the blue background metrics represent similarities (i.e.~the higher, the better). Notice also that \textit{all} these measures are plagued by the \textit{curse of dimensionality}.

\begin{figure}
    \centering
    \includegraphics[width=0.97 \textwidth]{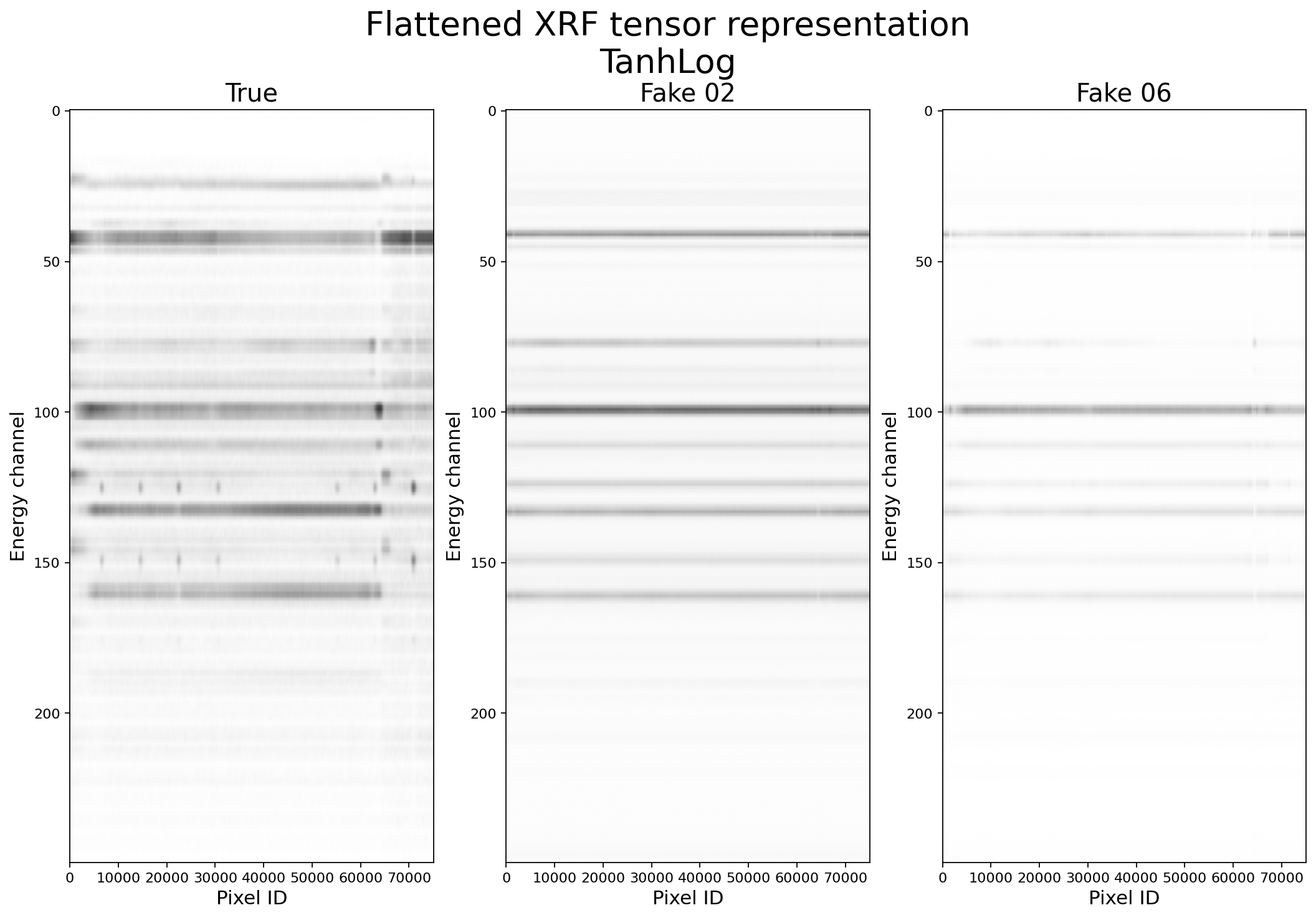}
    \caption{Flattened MA-XRF hypercubes comparison.}
    \label{fig:flattened_xrf_comparison}
\end{figure}

Secondly, we may flatten the XRF hypercube spatial dimensions, so that we have a 2D tensor (i.e.~a matrix) of shape $(H\cdot W, E)$, where $H$ is the height, $W$ the width, and $E$ the number of energy channels. The visual comparison of those tensors are reported in Figure \ref{fig:flattened_xrf_comparison}.

We have reported the binary map of the data cubes, after having applied a simple transform to enhance the visualisation of low peaks, i.e. we have mapped $x \mapsto \tanh( \log (1 + x) )$.

We can then use Structural Similarity Index Measure (SSIM) \cite{SSIM1284395}, and its multi-scale version (MS-SSIM) \cite{MSSSIM1292216}, to compute the distances among those. We also generate a random poissonian noise tensor (with the appropriate number of noise counts per pixel) as a reference measure. The results are reporte in Table \ref{tab:flattened_xrf_comparison_res}.

\begin{table}[h]
    \caption{SSIM and MS-SSIM Scores for varius flattened datacubes comparison. }
    \centering
    \label{tab:flattened_xrf_comparison_res}
    \begin{adjustbox}{max width=\textwidth}
        \begin{tabular}{l|cc|c|c}
         &
          \cellcolor[HTML]{9AFF99}True vs Fake 0.2 &
          \cellcolor[HTML]{9AFF99}True vs Fake 0.6 &
          \cellcolor[HTML]{FFFFC7}Fake 0.2 vs Fake 0.6 &
          \cellcolor[HTML]{FFCCC9}True vs Poisson Noise \\ \hline
        \cellcolor[HTML]{ECF4FF}SSIM &
          0.3491 &
          0.3317 &
          0.4865 &
          0.0878 \\
        \cellcolor[HTML]{ECF4FF}MS\_SSIM &
          0.4390 &
          0.3697 &
          0.7130 &
          0.1597
        \end{tabular}
    \end{adjustbox}
\end{table}

Nevertheless, the purpose of this section was merely to illustrate the potentiality of the \texttt{ganX} package. 

\section{Conclusion}

In this brief paper we have presented \texttt{ganX}, a Python package offering methods to generate artificial new MA-XRF data cubes out of an RGB image. This package may be helpful in the Syntethic dataset generation for applications of statistical and deep learning methods in the field of MA-XRF analysis, especially in the field of Cultural Heritage. 

The next steps of the ganX package project will be to extend the flexibility of the package, by adding additional features, like to possibility to generate MA-XRF data cubes without a fixed amount of signal per-pixel, but allowing for, e.g.~luminosity-based number of counts. Also, we are eploring the possibility of adding Poissonian noise, as well as other noise origin. 

Furthermore, we would like to add the actual "GAN" part to \texttt{ganX}, i.e.~creating a Generative Adversarial Network to generate the MA-XRF data cubes out of gaussian noise. 

\section*{Acknowledgments}
The present work has been partially funded by the European Commission within the Framework Programme Horizon 2020 with the project 4CH (GA n.101004468 – 4CH) and by the project \textit{AIRES–CH - Artificial Intelligence for digital REStoration of Cultural Heritage} jointly funded by Tuscany Region (Progetto Giovani Sì) and INFN. 

We thank the authors of \cite{heritage2020103} for sharing the raw data used in Section \ref{sec:simple_example}. The analysis conducted in the aforementioned paper on the Anphitionary M belonging to the collection of San Giorgio Maggiore  was funded by the Abbey of San Giorgio Maggiore. One of the author of \cite{heritage2020103} (A.M.) research fellowship, during which that work was undertaken, was funded by the Zeno-Karl
Schindler Foundation. In particular, we would like to thank  A.~Mazzinghi and C.~Ruberto for useful discussions.

\appendix
\section{Metrics}\label{app:metrics}

Here we report the definitions of the metrics used in Section \ref{sec:simple_example}. Notice that the histograms appearing in \ref{fig:hists} are normalised to 1, i.e.~can be seen as discrete probability distributions. 

\paragraph{Correlation Distance:} The \textit{correlation distance} between two 1D arrays $u,v$ is defined as
\begin{equation}
    \textsc{correlation}(u,v) = 1 - \frac{u \cdot v}{\norm{(u - \bar u)}_2 \,\norm{(v - \bar v)}_2 } \,,
\end{equation}
where $\bar{u}, \bar{v}$ are the mean elements of $u,v$.

\paragraph{Cosine Distance:} The \textit{cosine distance} between two 1D arrays $u,v$ is defined as
\begin{equation}
    \textsc{cosine}(u,v) = 1 - \frac{u \cdot v}{\norm{u}_2 \,\norm{v}_2 } \,.    
\end{equation}
The cosine distance is related to the angle $\theta$ between the vectors $u,v \in \mathbb{R}^N$, i.e.
\begin{equation}
    \cos \theta = \frac{u \cdot v}{\norm{u}_2 \,\norm{v}_2 } = 1 - \textsc{cosine}(u,v)  \,.
\end{equation}

\paragraph{Jensen - Shannon:} The \textit{Jensen-Shannon distance} between two normalised 1D arrays $u,v$ is defined as
\begin{equation}
    \textsc{JensenShannon}(u,v) = \sqrt{
    \frac{D_{KL}(u || m) + D_{KL}(v || m)}{2}
    }\,.,
\end{equation}
where $m$ is the pointwise mean of $u$ and $v$ and $D_{KL}$ is the \textit{Kullback-Leibler} divergence (or relative entropy), i.e.
\begin{equation}
    D_{KL}(P || Q) = \sum_{n=1}^{N} P(n) \log \frac{P(n)}{Q(n)} \,.
\end{equation}

\paragraph{Chebyshev:} The \textit{Chebyshev distance}  between two 1D arrays $u,v$ is defined as
\begin{equation}
    \textsc{Chebyshev} (u,v) = \max_i \left| u_i - v_i \right| \,.
\end{equation}

\paragraph{Bray-Curtis:} The \textit{Bray-Curtis} dissimilarity  between two 1D arrays $u,v$ is defined as
\begin{equation}
    \textsc{BrayCurtis} (u,v) = \sum_{n=1}^N \frac{ \left| u_n - v_n \right|}{ \left| u_n + v_n \right|} \,.
\end{equation}

\paragraph{Chi-quare} The \textit{Chi-square} distance  between two 1D arrays $u,v$ is defined as
\begin{equation}
    \chi^2 (u,v) = \frac{1}{2} \, \sum_{n=1}^N \frac{ \left( u_n - v_n \right)^2}{ \left| u_n + v_n \right|} \,.
\end{equation}

\paragraph{Bhattacharyya:} The \textit{Bhattacharyya coefficient} \textsc{BC} between two 1D arrays $u,v$ is defined as
\begin{equation}
    \textsc{BC} (u,v) =  \sum_{n=1}^{N} \sqrt{ u_n  v_n  } \,.
\end{equation}
The Bhattacharyya coefficient can be also used to compute the \textit{Bhattacharyya distance} $\textsc{D}_B$
\begin{equation}
    \textsc{D}_B ( u, v) = -\log \textsc{BC} (u,v) \,,
\end{equation}
as well as the \textit{Hellinger distance} $H$
\begin{equation}
    H(u,v) = \sqrt{1- \textsc{BC} (u,v) } \,.
\end{equation}

\bibliographystyle{unsrt}  
\bibliography{references}

\end{document}